\documentclass[twocolumn,superscriptaddress]{revtex4-1}%linenumbers,

\usepackage{graphicx}
\usepackage{float}
\usepackage{epstopdf}
\usepackage{amsmath}
\usepackage{amssymb}
\usepackage{mdframed}
\usepackage[utf8]{inputenc}
\usepackage[dvipsnames]{xcolor}
\usepackage{tikz}

\usetikzlibrary{automata,positioning,arrows.meta}
\usepackage[caption = false]{subfig}

\usetikzlibrary{automata,positioning,arrows.meta}
\usepackage[caption = false]{subfig}
\makeatletter

\newcommand{\Rmnum}[1]{\expandafter\@slowromancap\romannumeral #1@}

\makeatother

\begin{document}

\title{{Low complexity, low probability patterns and consequences \\for algorithmic probability applications}}

\author{Mohamed Alaskandarani}
\affiliation{Centre for Applied Mathematics and Bioinformatics (CAMB),
Department of Mathematics and Natural Sciences, 
Gulf University for Science and Technology, Kuwait}

\author{Kamaludin Dingle}
\email{Correspondence: dinglek@caltech.edu}
\affiliation{Centre for Applied Mathematics and Bioinformatics (CAMB),
Department of Mathematics and Natural Sciences, 
Gulf University for Science and Technology, Kuwait}
\affiliation{Department of Chemical Engineering and Biotechnology, Cambridge University, UK}
\affiliation{Department of Computing and Mathematical Sciences, Caltech, USA}

\date{\today}

\begin{abstract}
\noindent
Developing new ways to estimate probabilities can be valuable for science, statistics, and engineering. By considering the information content of different output patterns, recent work invoking algorithmic information theory has shown that a priori probability predictions based on pattern complexities can be made in a broad class of input-output maps. These algorithmic probability predictions do not depend on a detailed knowledge of how output patterns were produced, or historical statistical data. Although quantitatively fairly accurate, a main weakness of these predictions is that they are given as an upper bound on the probability of a pattern, but many low complexity, low probability patterns occur, for which the upper bound has little predictive value. Here we study this low complexity, low probability phenomenon by looking at example maps, namely a finite state transducer, natural time series data, RNA molecule structures, and polynomial curves. Some mechanisms causing low complexity, low probability behaviour are identified, and we argue this behaviour should be assumed as a default in the real world algorithmic probability studies. Additionally, we examine some applications of algorithmic probability and discuss some implications of low complexity, low probability patterns for several research areas including simplicity in physics and biology, a priori probability predictions, Solomonoff induction and Occam's razor, machine learning, and password guessing.

\vspace{0.2cm}
\end{abstract}

\maketitle
\twocolumngrid

\section{Introduction}

Many problems in science, statistics, and engineering revolve around estimating probabilities of events. This is especially true in the current climate, where machine learning and data science are enjoying broad applications. Hence developing new methods for calculating, bounding, or predicting probabilities is valuable. One direction for such predictions is in theoretical computer science where \emph{algorithmic information theory} \cite{solomonoff1960preliminary,kolmogorov1965three,chaitin1975theory,li2008introduction} (AIT) provides a theoretical framework for studying randomness,  probability, and complexity. The central quantity of AIT is \emph{Kolmogorov complexity}, $K(x)$, which measures the complexity of an individual object or pattern $x$ via the amount of information required to describe or generate $x$. The object could be a binary string, integer, graph, or indeed anything that can be represented as a binary string. The AIT coding theorem \cite{levin1974laws} establishes a fundamental connection between complexity and probability predictions in very general settings. It states that the chances that some output $x$ is produced via some generic computation mechanism  is directly related to the complexity of $x$.  More formally, the coding theorem states that $P(x)\sim 2^{-K(x)}$ where $P(x)$ is the probability that an output $x$ is generated by a (prefix optimal) universal Turing machine fed with a random binary program input. $P(x)$ is known as the \emph{algorithmic probability} of $x$, and the associated probability distribution is known as the \emph{universal distribution} \cite{solomonoff2003kolmogorov}.  

\begin{figure*}%[h]
\subfloat[]{\includegraphics[width=9cm]{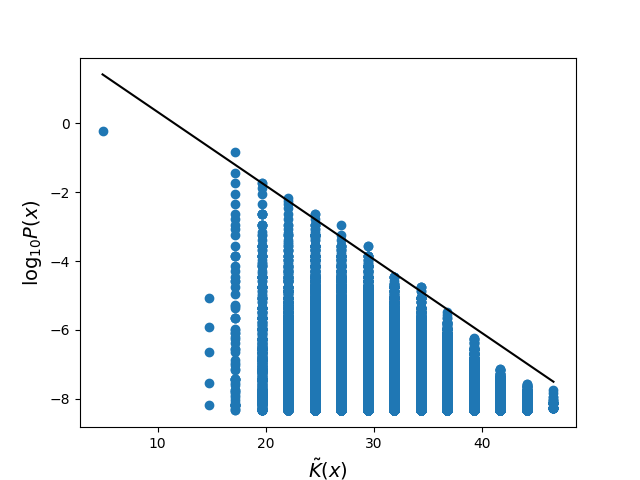}} %,trim={0 0 0 1cm},clip
\subfloat[]{\includegraphics[width=9cm]{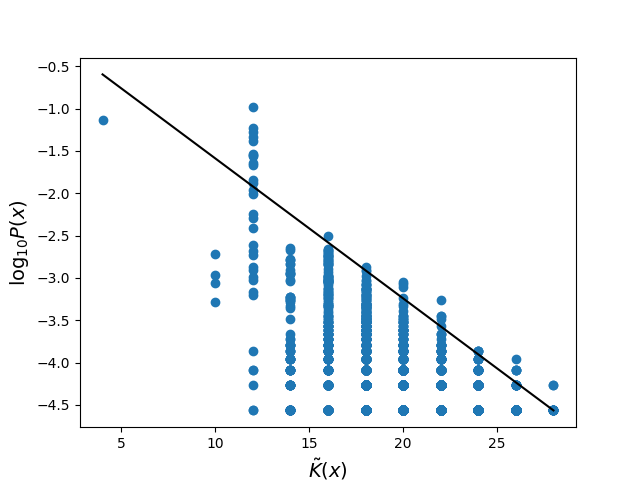}} \\
\subfloat[]{\includegraphics[width=9cm]{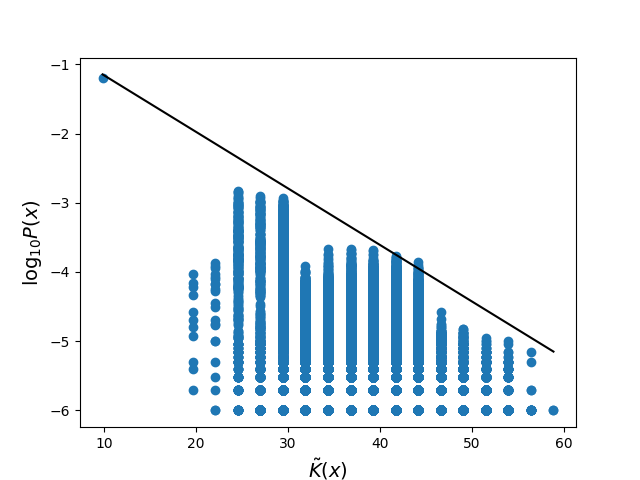}} 
\subfloat[]{\includegraphics[width=9cm]{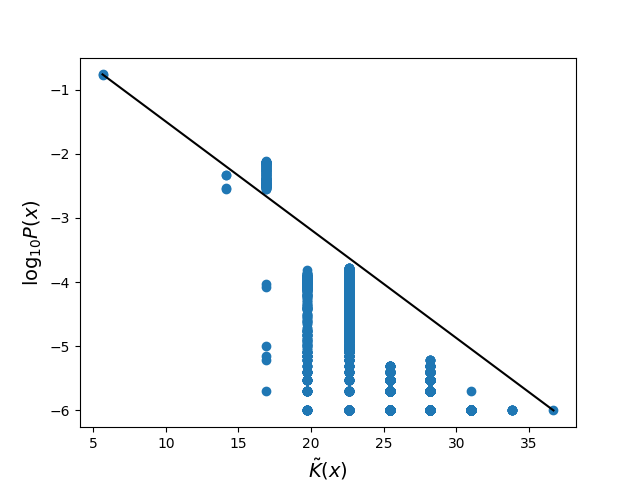}} 
\caption{\textbf{Simplicity bias plots with low complexity, low probability outputs.} Blue dots represent different binary string output patterns. The black line is the upper bound of Eq.\ (\ref{eq:SB}). Probabilities $P(x)$ are the fraction of random inputs that map to output $x$. $\tilde{K}(x)$ are estimated complexity values for each output $x$.  There are many data points far below the upper bound; these points are low Kolmogorov complexity, low probability (LKLP) output patterns. The maps studied are: (a) A finite state transducer (FST) with data taken from ref.\ \cite{dingle2020generic}; (b) time series patterns from the World Bank Open Data, with data taken from \cite{dingle2022note}; (c) computationally predicted RNA secondary structures from 1 million randomly sampled input sequences; and (d) discretised polynomials curves generated with $x\in$(0,1) and random coefficients, data taken from \cite{dingle2018input}.
}
\label{fig:SB}
\end{figure*}

Algorithmic probability and  AIT results are typically difficult to apply in real-world settings due to the fact that $K(x)$ is uncomputable, the theorems assume the presence of universal Turing machines (UTMs), and the results are asymptotic and stated with accuracy to within an unknown constant. Despite these theoretical difficulties, in practice many successful applications of AIT have been made based, for example in bioinformatics \cite{cilibrasi2005clustering,ferragina2007compression}, physics \cite{avinery2019universal}, signal denoising \cite{vitanyi2013similarity}, among many other applications \cite{li2008introduction}. Mostly these applications use standard compression algorithms to approximate $K(x)$, sometimes combined with various forms of theorem approximation.  Algorithmic probability estimates have been made numerically by random sampling \cite{legg2011approximation} and enumeration \cite{delahaye2012numerical,soler2014calculating,zenil2019coding} of computer programs.

From a very different perspective, algorithmic probability estimates have also been made via deriving a weaker form of the coding theorem, applicable in real-world contexts \cite{dingle2018input}, taking the form of an upper bound. This weaker form upper bound was applied in a range of input-output maps to make a priori predictions of the probability of different shapes and patterns, such as the probability of different RNA shapes appearing on a random choice of genetic sequence, or the probability of differential equation solution profile shapes, on random choice of input parameters, and several other examples \cite{dingle2018input,dingle2020generic}. Surprisingly, it was found that probability estimates could be made directly from the complexities of the shapes themselves, without recourse to the details of the map or reference to how the shapes were generated. The authors of \cite{dingle2018input} termed this phenomenon of an inverse relation between complexity and probability \emph{simplicity bias} (SB). One important drawback of this work is that only an upper bound prediction was made, rather than a direct probability value prediction. In contrast to original coding theorem, in practice it has been observed that many simple patterns have low probability  \cite{dingle2018input}. Such low Kolmogorov complexity, low probability (LKLP) outputs present a weakness in the predictive ability of the upper bound, because for these outputs their complexity and probability values are largely disconnected and hence predicting one from the other is more challenging. Understanding the causes and properties of LKLP outputs may help to improve the accuracy of applications of algorithmic probability, such as better a priori probability predictions and induction. In this work we investigate LKLP behaviour, and its implications for applied algorithmic probability via some examples from previously published works.

\section{Background theory}
Some brief accessible background theory is given here; more formal and detailed presentations are available in, eg refs.\ \cite{li2008introduction,calude2002information,gacs1988lecture}.
A universal Turing machine (UTM) \cite{turing1936computable} is an abstract general computing device which can simulate any other Turing machine. A UTM has the highest computational capacity, and can implement any conceivable algorithm which could in principle be run on a computer. A programming language is called \emph{Turing complete} (or computationally universal) if the language is sufficiently expressive to be able to simulate a UTM, and therefore implement any algorithm. Common languages like Python, C, FORTRAN, etc are Turing complete. If a function can be computed by a finite mechanical procedure, then it a computable function. For computable functions, all inputs or programs eventually halt, and cannot (for example) run on forever in an infinite loop. Common functions like polynomials, exponentials, trigonomentric functions, etc are computable. 

The Kolmogorov complexity $K_U(x)$ of a string $x$ with respect to $U$,  is defined \cite{solomonoff1960preliminary,kolmogorov1965three,chaitin1975theory} as
\begin{equation}
K_U(x) = \min_{p}\{|p|: U(p)=x\}
\end{equation}
where $p$ is a binary program for a prefix optimal UTM $U$, and $|p|$ indicates the length of the binary program $p$ in bits.  Due to the invariance theorem \cite{li2008introduction} for any two optimal UTMs $U$ and $V$, $K_U(x) = K_V(x)+O(1)$ so that the complexity of $x$ is independent of the machine, up to additive constants. Hence  we conventionally drop the subscript $U$ in $K_U(x)$, and speak of `the' Kolmogorov complexity $K(x)$. Informally, $K(x)$ can be defined as the length of a shortest program that produces $x$, or simply as the size in bits of the compressed version of $x$. If $x$ contains repeating patterns like $x=1010101010101010$ then it is easy to compress, and hence $K(x)$ will be small. On the other hand, a randomly generated bit string of length $n$ is highly unlikely to contain any significant patterns, and hence can only be described via specifying each bit separately without any compression, so that $K(x)\approx n$ bits. Other names for $K(x)$ are \emph{descriptional complexity}, \emph{algorithmic complexity}, and \emph{program-size complexity}. Fundamentally, $K(x)$ measures the amount of information to describe or generate $x$ precisely and unambiguously. Note that Shannon information and Kolmogorov complexity are closely related to each other \cite{grunwald2004shannon}, but also differ fundamentally because Shannon information quantifies the information content of a random source, while Kolmogorov complexity quantifies the information of individual sequences or objects.  

Solomonoff invented algorithmic probability \cite{solomonoff2009algorithmic}, but it was later formalised and extended by Levin \cite{levin1974laws}, who proved the AIT \emph{coding theorem} in 1974 which states that
\begin{equation}
P(x) = 2^{-K(x)+O(1)}\label{eq:CD}
\end{equation}
where $P(x)$ is the probability that UTM $U$ generates output string $x$ on being fed random bits as a program (again we have dropped the subscript $U$). Thus, high complexity outputs have exponentially low probability, and simple outputs must have high probability. %This is a profound result which links notions of data compression and probability in a direct way. %$P(x)$ is also known as the \emph{algorithmic probability} of $x$.

Coding theorem-like behaviour in real-world input-output maps was studied, leading to the observation of a phenomenon called \emph{simplicity bias} (SB) \cite{dingle2018input}. SB is captured mathematically as
\begin{equation}
P(x)\leq 2^{-a\tilde{K}(x)-b}\label{eq:SB}
\end{equation}
where $P(x)$ is the (computable) probability of observing output $x$ on random choice of inputs, and $\tilde{K}(x)$ is the estimated Kolmogorov complexity of the output $x$: complex outputs from input-output maps have lower probabilities, and high probability outputs are simpler. The constants $a>0$ can often be estimated without recourse to sampling, but just by knowing or estimating the total number of different possible outputs \cite{dingle2018input}. Using $b=0$ is the default guess for this constant, but it can also be fit to the data via partial sampling. 
%The ways in which SB differs from Levin's coding theorem include that it does not assume UTMs, uses approximations of complexities, and for many outputs $P(x)\ll 2^{-K(x)}$. Hence the abundance of LKLP outputs is a signature of SB. 

A complete understanding of exactly which maps will show SB has not been developed, but  SB is expected to appear in many input-output maps, under fairly general conditions. Importantly, the map should be `simple' (technically of $O(1)$ complexity) to prevent the map itself from dominating over inputs in defining output patterns. If an arbitrarily complex map was permitted, outputs could have arbitrary complexities and probabilities, and thereby remove any connection between probability and complexity.  Strong bias is a prerequisite for SB, typically the relative value of the largest  and smallest probabilities should be more than the value of the number of outputs, ie for $N_O$ outputs, $\max(P(x))/\min(P(x)) > N_O$.

\section{Results}

\subsection{Low complexity, low probability behaviour is very common}

Based on numerical experiments using the upper bound of Eq.\ (\ref{eq:SB}), Dingle et al \cite{dingle2018input} reported that many outputs $x$ have probability values far below their predicted upper bounds, ie $P(x)\ll 2^{-a\tilde{K}(x)-b}$ for many $x$. %Indeed, it was shown \cite{dingle2018input,dingle2020generic} numerically that typically many, perhaps most, of the outputs in the maps studied appeared with probabilities far below their respective upper bounds. 
This observation was not due to the bound being trivially loose, because it is a tight bound for many outputs. Significantly, the small fraction of outputs which were close to the upper bound absorbed most of the probability mass, or in other words most of the inputs map to outputs for which the bound is tight.  As a consequence it was shown \cite{dingle2018input,dingle2020generic} analytically and numerically that for an output $x$ generated by a random input, with high probability $P(x)\approx 2^{-a\tilde{K}(x)-b}$. 

To illustrate the LKLP phenomenon, in Figure \ref{fig:SB} we show four probability-complexity plots. The probabilities $P(x)$ are calculated as the fraction of random inputs which produce output $x$. The complexity $\tilde{K}(x)$ denotes an estimate of the Kolmogorov complexity of each output, using a slightly adapted \cite{dingle2018input} version of the famous Lempel-Ziv \cite{lempel1976complexity} 1976 complexity measure. The black lines are fitted upper-bounds, depicting the SB upper bound of Eq.\ (\ref{eq:SB}). All four maps have been studied in earlier works, and are: (a) a finite state transducer (FST), which is a simple generic model of computation, but unlike a UTM has a very limited computational capacity (being the lowest on the Chomsky hierarchy). Here the outputs are length 30 bits, and were obtained from thorough sampling of binary string inputs, and the data was taken from ref.\  \cite{dingle2020generic}. (b) Time series data taken from the World Bank Open Data project (\texttt{https://data.worldbank.org}), which have been discretised to binary strings (of length 16 bits) while studying SB \cite{dingle2022note}. (c) Computationally predicted \cite{lorenz2011viennarna} RNA secondary structures obtained from randomly sampling 1 million sequences of length $L=35$ nucleotides. Following the protocol of ref.\ \cite{dingle2018input} in which SB in computationally generated RNA structures was first reported, predicted dot-bracket structures were converted to binary strings and complexity values were thereby estimated. (d) Polynomial curves $y=\sum_{i=1}^{14} \alpha_i x^i$ with Gaussian random coefficients $\alpha_i$ and $x\in(0,1)$, with data from \cite{dingle2018input}. The curves were discretised to binary strings by the up/down method \cite{hansen2001model,willbrand2005identifying}. More details of these maps and relevant analyses can be found in the original cited papers.

It is apparent that in each panel of Figure \ref{fig:SB} the data points show a similar `triangle' shape, with some points closely following the upper bound (black line), but at the same time for each complexity value a large variation in probability values is observed, inferred from the many points far below the bound. Occupying the bottom left corner of the `triangles' are the outputs which exhibit the strongest LKLP behaviour, because they have very low probability, but at the same time have relatively low complexity values. This LKLP `triangle' depicted in all maps of Figure \ref{fig:SB} appears in essentially all examples of SB studied so far, including in biology \cite{johnston2022symmetry}, machine learning \cite{valle2018deep}, and other contexts \cite{dingle2018input} (Supp.\ Info.).

In Figure \ref{fig:SB} and in most other SB examples cited, the previously described complexity estimator $\tilde{K}(x)$ based on the Lempel-Ziv complexity measure was used. However,  LKLP outputs have also been observed when using other complexity measures, such as those used in protein quaternary structures and polyominos \cite{johnston2022symmetry}, so LKLP outputs are not merely an artefact of the Lempel-Ziv complexity measure. From another angle, nearly all the cited complexity-probability plots showing SB were generated via random sampling of inputs, and so it might be suggested that with more sampling, or full enumeration of inputs, LKLP outputs would disappear or at least become much less pronounced. However, this suggestion is countered by the observation that LKLP behaviour is also observed in maps for which probabilities were directly calculated by enumerating every possible input, for example L-systems \cite{dingle2018input} and RNA length $L=15$ nucleotides \cite{dingle2020generic}. Perhaps the only clear example of a map which does not show strong LKLP behaviour is in a small polyomino system using path complexity (see Fig 4.2(a) in Chapter 4 of ref.\ \cite{dingle2014probabilistic}). %In this polyomino system, there were relatively few outputs, but with a large range of complexities, which may be res.

To avoid confusion, we stress that LKLP behaviour is not due to a failing of complexity measures to detect patterns. If, say, the Lempel-Ziv measure failed to detect an important pattern and thereby assigned a high complexity to an output which is actually simple, then the opposite behaviour would be observed in the complexity-probability plots, where some outputs would be far \emph{above} the upper bound, not far below the bound. 

A well known limitation of the original coding theorem is choice of UTM, the problem being that algorithmic probability estimates from different machines will differ \cite{leike2015bad}, and there is no known way to remove this dependence \cite{muller2010stationary}. However, due to the invariance theorem, this machine dependence is only a problem for (typically small) outputs with low complexities, for which $O(1)$ terms due to translating between UTMs may dominate, but this is not a problem for large complexity strings. This machine dependence limitation may not be of great concern perhaps, because many real-world phenomenon are highly complex. In contrast, LKLP outputs do not disappear for large complexity outputs, and so LKLP can be seen as a more serious problem for applied algorithmic probability than machine dependence.

Finally, in many maps which exhibit SB, some patterns are actually impossible to make, such that $P(x)=0$, so this could be viewed as an extreme form of LKLP behaviour.

\subsection{Estimating which of two outputs is more likely}\label{whichishigher}

Another way to examine and quantify LKLP behaviour is in terms of predicting which of two outputs has higher probability. In the absence of LKLP outputs, if $K(x)<K(y)$ for two strings $x$ and $y$ then it follows from Eq.\ (\ref{eq:CD}) that $P(x)>P(y)$, ignoring $O(1)$ terms. With LKLP behaviour, the direct connection between probability and complexity is disrupted, and $K(x)<K(y)$ does not imply that $P(x)>P(y)$ necessarily. Note that predicting which of two strings has higher probability does not depend on estimating or fitting the parameters $a$ and $b$ in Eq.\ (\ref{eq:SB}); only the complexity values are required. This kind of calculation was done earlier by \cite{dingle2018input,dingle2022note}. 

The manner in which the strings $x$ and $y$ are chosen also affects the strength of the connection between complexity and probability. If $x$ and $y$ result from randomly chosen inputs, then they will be sampled with weights according to their probabilities, and in this case $P(x)$$\approx$$2^{-a\tilde{K}(x)-b}$ and $P(y)$$\approx$$ 2^{-a\tilde{K}(y)-b}$ \cite{dingle2020generic}. In this probability-weighted sampling scenario, for randomly generated outputs, the probability-complexity connection is quite strong. Hence we should expect to be able to predict whether $P(x)>P(y)$ or $P(x)<P(y)$ just from the complexity values, with quite high accuracy. If instead $x$ and $y$ are chosen uniformly randomly from the full set of possible outputs, then we can expect the probability-complexity connection to be less strong, and predicting whether $P(x)>P(y)$ or $P(x)<P(y)$ just from the complexity values will be less accurate. %As a side comment, we can also imagine a third possibility in which the full set of possible outputs for the map was not known in advance. In this case which is if $x$ and $y$ were randomly chosen from $\{0,1\}^n$ where $n=|x|=|y|$, in this case the probability-complexity connection may be even weaker, because potentially many string in $\{0,1\}^n$ cannot even be made by the map, and hence they would have $P(x)=0$ and or $P(y)=0$.

Here we computationally study the question of which of two strings has higher probability, both using the randomly generated pairs $x,y$ and also uniformly sampled $x,y$. The protocol we employ is: predict $P(x)>P(y)$ if $K(x)<K(y)$, $P(x)<P(y)$ if $K(x)>K(y)$, and we randomly guess which has higher probability if $K(x)=K(y)$. Note that there will be relatively few unique complexity values (eg only $\sim n$ values for strings of length $n$), and hence it happens quite commonly that two random strings have the same complexity, while it is relatively rare for two strings to have exactly the same probability value. 

For both probability-weighted sampling and uniform sampling, the protocol just described has a null prediction accuracy level of 50\% (assuming a null hypothesis that output complexity has no predictive value). The calculated accuracy values (after 10,000 sampled pairs) are: FST probability-weighted sampling achieved an accuracy of 79\% and uniform sampling an accuracy of 63\%; time series 81\% and 80\%; RNA 78\% and 71\%; polynomial 82\% and 73\%. It is quite striking that high levels of accuracy can be achieved just by complexity estimates, which does not even require estimating the values of $a$ or $b$. As expected, the accuracy values for probability-weighted sampling are higher, while substantially lower for uniform sampling, but still substantially above the 50\% null model mark. Overall, the effect of LKLP outputs it clear: while some \% accuracy values are (surprisingly) high, they are still  substantially below 100\% which is what we would expect in the absence of LKLP outputs.

Note that true uniform sampling strictly requires knowing the full set of outputs, which might entail fully enumerating all inputs, or very thorough input sampling. In the four maps studied here, only the FST map was thoroughly sampled whereas the other maps were sampled partially (eg for RNA, only $10^6$ sampled inputs out of a possible $4^{30}\sim10^{18}$ RNA sequence inputs were made). Hence our stated accuracy values for `uniform sampling'  will be somewhat overestimates for RNA, time series, and polynomials, but for the FST map the accuracy value of 63\% is likely to be close to the true value. %The fact that the accuracy values are substantially above 50\% will be seen to be important when we look at induction below.

\begin{figure*}%[h]
\subfloat[]{\includegraphics[width=9cm]{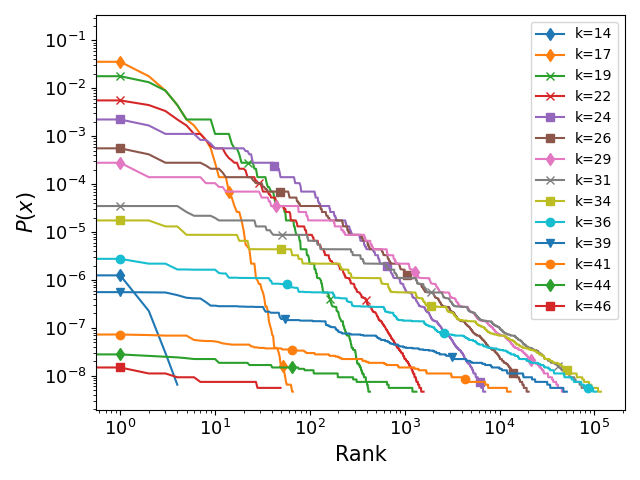}} %,trim={0 0 0 1cm},clip
\subfloat[]{\includegraphics[width=9cm]{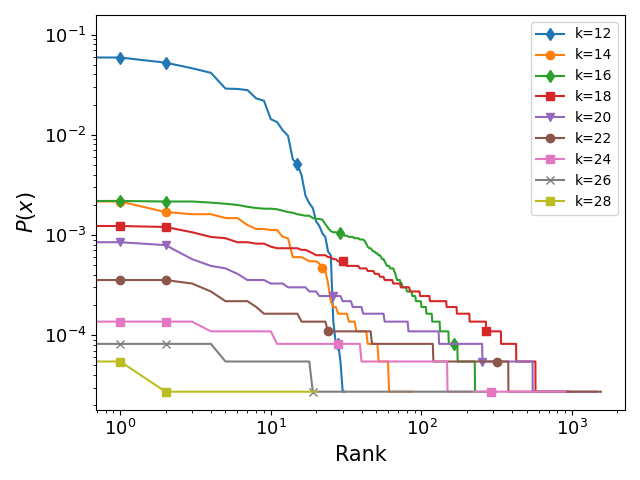}} \\
\subfloat[]{\includegraphics[width=9cm]{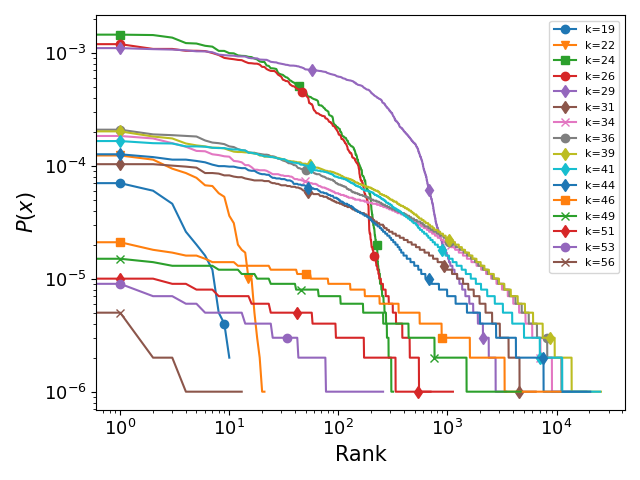}} 
\subfloat[]{\includegraphics[width=9cm]{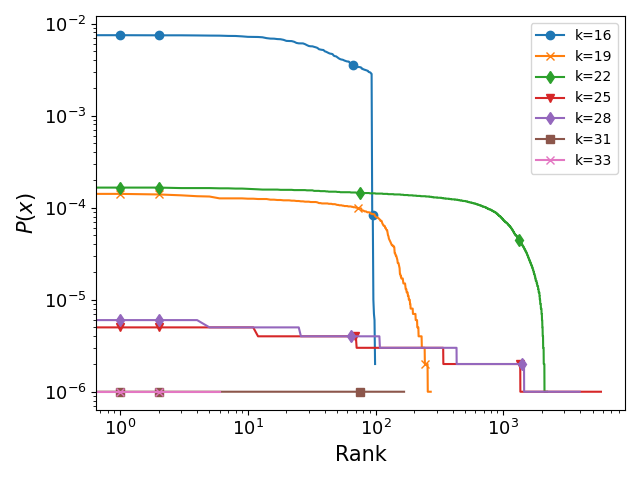}} 
\caption{\textbf{Rank plots of probabilities, separated by complexity values.} Each coloured line depicts the decay in probability of outputs with the same complexity value $k$. The complexity values $k$ are shown in the figure legends. The plots present the same data as in Figure \ref{fig:SB}, but now plotted as rank plots separated by complexity value. (a) FST; (b) time series; (c) RNA structures; and (d) random polynomials. In each case, a similar pattern of decay is observed, and almost all rank plots decay to the lowest probability for that map.}
\label{fig:rank} 
\end{figure*}

%migth want to remove
\vspace{1cm}

\subsection{Why do low complexity, low probability\\ outputs exist?}

We have seen LKLP outputs in essentially all the maps we have studied. Why are they so common? In the original AIT coding theorem given in Eq.\ (\ref{eq:CD}), there are no LKLP output patterns, except in the rather modest sense that the $O(1)$ term in the exponent allows for relatively small variation of probabilities within a constant, mathematically $|K(x)-\log_2(1/P(x))|<c$ for some constant $c$. The reason simple outputs cannot have very low probability in the original AIT coding theorem is that the theorem is based on random prefix codes run on a UTM. In a UTM an output $x$ always has a program $p$ of length $K(x)$ by definition of Kolmogorov complexity. The probability that the specific program $p$ appears as the first $p$ bits of the random binary program is $2^{-K(x)}$ and therefore $P(x)\geq 2^{-K(x)}$ implying that all simple strings must have high probability. Hence in this UTM setting, it is not possible to have LKLP outputs with $P(x)\ll 2^{-K(x)+O(1)}$.

Computable maps lack some computational power as compared to UTMs, which means that there are some algorithms which they cannot implement (for example, ones that don't halt).  As suggested earlier \cite{dingle2020generic} the weaker computational ability of these computable maps suggests that they may not be able to compute some output patterns which are nonetheless simple, or at least cannot compute them with efficient short programs. Hence we can expect some type of LKLP in computable maps, and the lower bound $P(x)\geq 2^{-K(x)}$ does not necessarily hold. %For example, RNA structures have geometric constraints based on chemistry preventing them from adopting certain shapes, or in other words computing certain patterns.  

Looking more closely at the lower bound vs.\ upper bound contrast, it appears that the upper bound on $P(x)$ is more fundamental. The justification is that if the upper bound was violated, and some complex outputs had high probability,  then it would be as if the map itself was creating information, which is not possible, assuming that the map is `simple' with $O(1)$ complexity. In contrast, no fundamental information theory violations arise if a simple output has a low probability, which helps to understand their common occurrence.

It is worth noting that LKLP behaviour is not actually a mathematical necessity for computable maps in general if a computable estimate of complexity is used. To see why, one could easily directly construct a map which has outputs with probability $2^{-\tilde{K}(x)}$, for some (computable) choice of $\tilde{K}(x)$ such as a Lempel-Ziv complexity measure. By construction, all probabilities would sit on the upper bound, with no LKLP outputs. A less contrived instance where LKLP behaviour may be less common is in maps for which inputs are organised into `constrained' and `unconstrained' regions \cite{greenbury2015organization}, which is similar to a Shannon-Fano coding. Having said that, even in this setting it is possible to have simple patterns with low probability.

In ref.\ \cite{dingle2020generic} it was suggested that LKLP outputs are intrinsically simple yet `hard' to make for the specific map, and it was shown analytically that such outputs coincide with those that can only be generated (via the map) with simple input. Indeed, a lower bound on $P(x)$ based on complexity inputs and outputs was derived.

Illustrating the `hard to make' argument in the present context, we can take time series as an example, for which we find that the bit strings $x$=1111111011111111 and $y$=0000000011111111 in Figure \ref{fig:SB}(b) were assigned the same low complexity value, and yet $y$ had a $\sim$$10^3$ fold higher probability than $x$, which was a LKLP output. Thinking about common time series patterns from everyday experience, we can see that $y$ is `easy' to make for a time series because the binary string pattern $y$ would result from essentially any gradually increasing series, such as linear or exponential growth, both of which are very common. In contrast, even though $x$ is simple --- just a string of 1's with a single 0 in the middle --- it would be `hard' to make such a string because in the discretisation process  a `1' denotes a series value above the mean value of the series, and 0 denotes below. Hence to generate $x$ would require one very low value in the middle of a stretch of high values, which is quite unlikely to occur in a time series. Given the way the series was discretised in terms of above/below the mean value of the string, we can expect that strings with an excess of 1's or 0's will have low probability, even if they are not complex. For RNA secondary structure a `hard to make' output was given in ref.\ \cite{dingle2020generic} where it was pointed out that the dot-bracket representation of an RNA structure such as (.(.(.(.(.(.(.(.(.(...).).).).).).).).).) would be thermodynamically unfavourable due to many lone chemical bonds, and therefore it would have very low probability, despite being a simple symmetric structure.
More generally, it is not hard to see that a given map will have certain biases towards or against certain patterns, and these map-specific biases can affect probabilities strongly, independently of pattern complexity. 

Another way that LKLP outputs can occur is if the complexity measure is too coarse, assigning too few complexity values. For a binary string of length $n$, there are around $n$ possible complexity values, so if a measure assigns much less than $n$ complexity values, then different complexity values will be combined together into one grouping, but their true complexities will actually be varied, and hence we can expect their probabilities to vary also, apparently yielding LKLP outputs.

%We do not claim to have an explanation at this point, but one possible direction to addressing this is to note that for a simple set, any extremes in probability must be associated to simple outputs: if any outputs have very high probability or a few outputs have very \emph{low} probability then these must be simple. To see why, consider a simple set ---  take the set of all binary strings of length $n$ --- with a simple $O(1)$ map used to assign probabilities to each output. Now rank the outputs by probability, highest to lowest. We can describe any output in the list by the index of its rank, so for example the highest probability output will have index $i=1$, and so it can be described in $K(1)+O(1)=O(1)$ bits, and so it must be very simple. Now imagine if there is one or a few outputs with exceptionally low probability, then these can also be easily described, due to being easily isolated from the majority of other strings, and so must be simple. From this reasoning, we can infer that low complexity strings are perhaps likely to occur with extreme probabilities, which might help to understand why LKLP outputs often have very low probability. This argument does not hold however if very many  This observation accords with the work 
The fact that the LKLP `triangle' is common to essentially all maps studied so far in the literature suggests there may be a general explanation. Burkat \& Dingle \cite{burkat2022note} studied SB in the Bernoulli process with parameter $\neq0.5$, and reported that both the single highest and single lowest probability outputs are also the simplest, and pointed out the cause of LKLP outputs in the Bernoulli map is easy to understand. As an extension, we propose here that if an output is (even roughly) made up of statistically independent parts, then the output may be approximated as a (biased) Bernoulli process, for which we expect LKLP behaviour \cite{burkat2022note}. As an example, for a time series of some fixed length $n$, the non-adjacent values are typically correlated only weakly, hence roughly independent, especially if separated by longer intervals. Even for RNA secondary structure for which there are definite correlations between distant parts of the sequence due to bonding interactions, but nonetheless much of the structure is roughly independent of other structure provided the sequence is not very short. This line of argument might help to explain some instances of LKLP outputs. 

\begin{figure*}%[h]
\subfloat[]{\includegraphics[width=9cm]{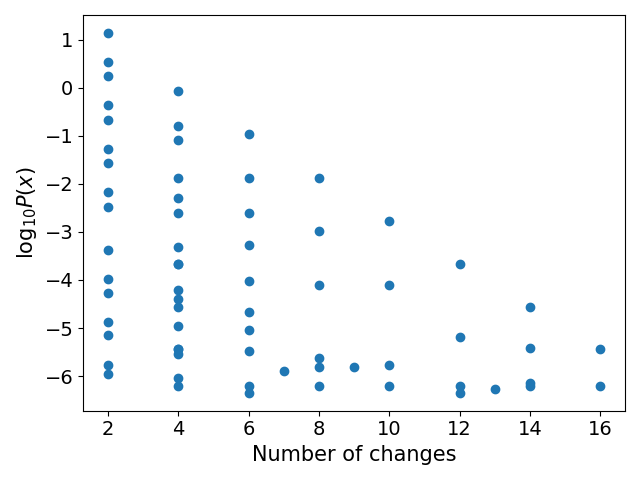}} %,trim={0 0 0 1cm},clip
\subfloat[]{\includegraphics[width=9cm]{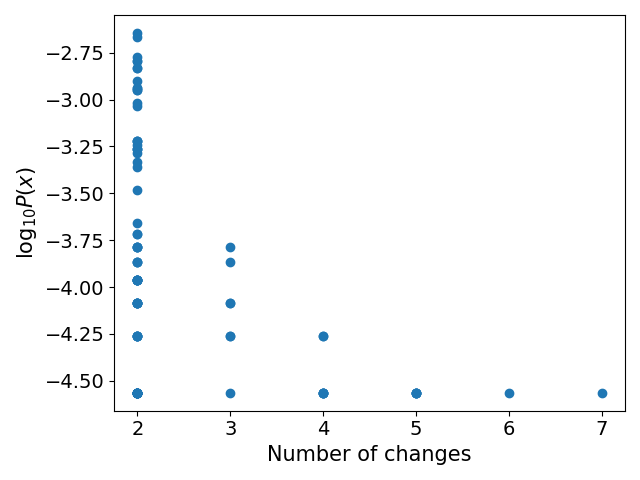}} \\
\subfloat[]{\includegraphics[width=9cm]{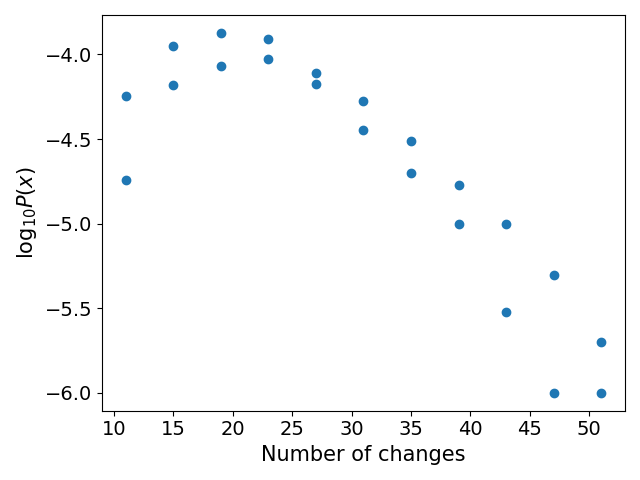}} 
\subfloat[]{\includegraphics[width=9cm]{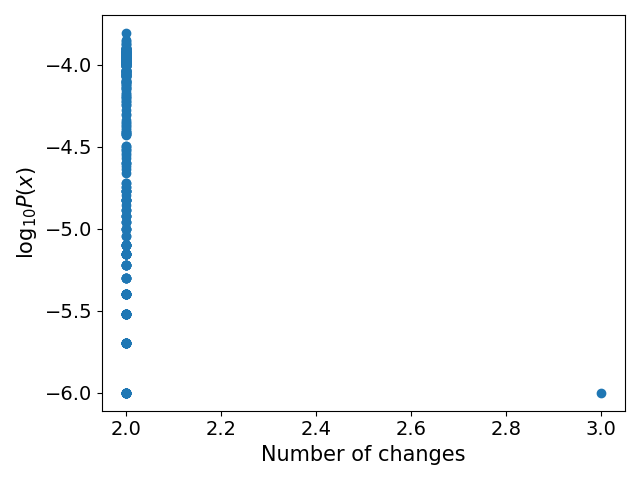}} 
\caption{\textbf{Number of 0/1 of changes in an output string vs the output probability, for just one complexity value $k$.} (a) For the FST map, there is a clear decay in probability with increasing numbers of 0/1 changes, using complexity $k=$ 17;  (b) similarly for the time series data, there is a clear decay, using $k=$ 14; (c) For the RNA there is a strong decay after an initial increase, using $k=$ 22; (d) no trend for the polynomials data, using $k=$ 19.}
\label{fig:changes}
\end{figure*}

\begin{figure*}%[h]
\subfloat[]{\includegraphics[width=9cm]{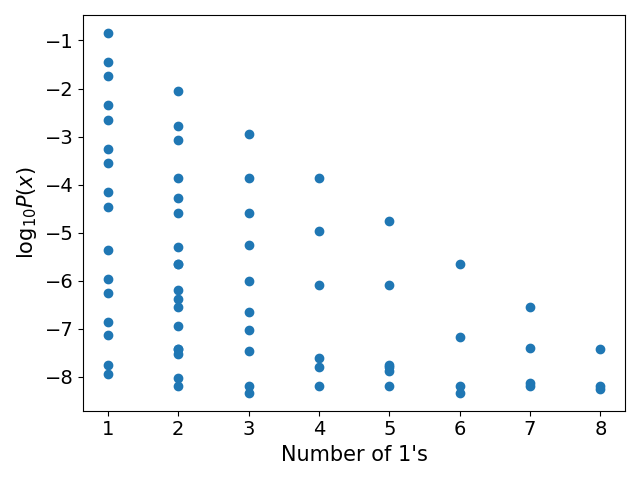}} %,trim={0 0 0 1cm},clip
\subfloat[]{\includegraphics[width=9cm]{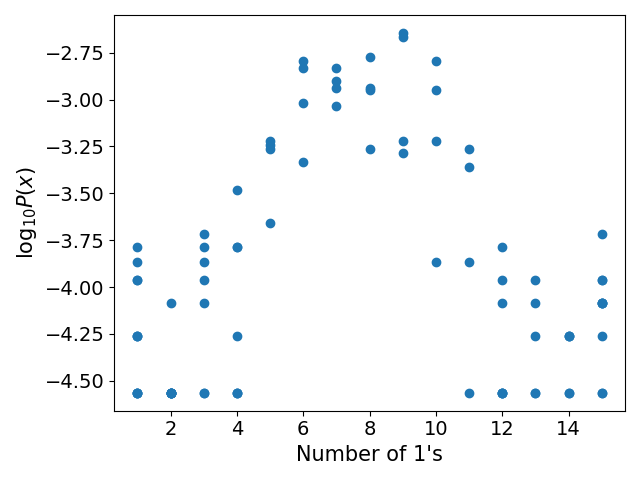}} \\
\subfloat[]{\includegraphics[width=9cm]{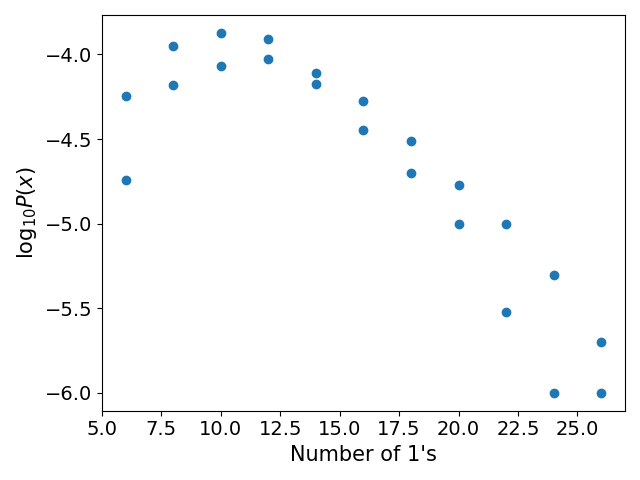}} 
\subfloat[]{\includegraphics[width=9cm]{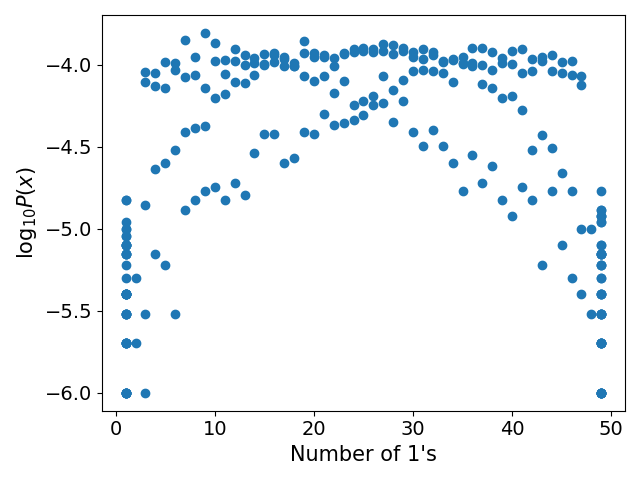}} 
\caption{\textbf{The number of 1's in an output string vs the string's probability.} For each example map a plot is given for just one complexity value $k$, the second simplest, chosen to aid illustration. (a) The FST example shows a linear trend in log probability with increasing number of 1's, using complexity $k=$17; (b) Large and small numbers of 1's are not favoured in the time series, using $k=$14; (c) RNA shows an overall negative relation, using $k=$22; (d) polynomials, similar to time series, large and small numbers of 1's are not favoured, using $k=$19.}
\label{fig:numb_ones}
\end{figure*}

\subsection{Rank plots of probability values, separated by complexity}

We now study the distribution of LKLP probabilities for the four maps shown in Figure \ref{fig:SB} above.  Within Figure \ref{fig:SB}, LKLP outputs appear as a `column' of overlapping blue dots stretching down from the black line upper bound, but the distribution of those LKLP probability values is not easy to discern. Instead in Figure \ref{fig:rank} for each of those maps, a loglog plot is given  showing the rank and probability of each output, coloured separately by complexity value $k$, so that the distribution is more easily visualised.

There are some common patterns to these disparate rank plots: In nearly all cases the rank plots for all complexity values decay to the same lowest probability, eg for the FST data in Figure \ref{fig:rank}(a) all the complexity values from $k=14$ to $k=46$ decay to a similar low probability value of close to $10^{-8}$. It is not clear exactly why this occurs with such consistency. This needs to be explored in future work.

Another observation is that the probability decay follows a similar form, with the lowest complexity curves starting at a high probability and quickly decaying, and the higher complexity curves starting lower probability and slowly decaying. This observation can be rationalised by noting that for low complexities, there are exponentially few possible outputs (ie few simple patterns), yet some of these have exponentially high probability as expected from SB. As the complexity increases, it is well known from AIT that there are exponentially more possible patterns for a given complexity value, yet the maximum probability decays exponentially (consistent with the upper bound of Eq.\ \ref{eq:SB}), which explains why the rank plots become wider and less tall for higher complexities.

Interestingly, in the FST map, which has the cleanest data, the profile of the different curves appear to depict a straight line decay on the loglog plot, suggesting some kind of Zipf's power law upper bound on the rank decay profiles. Although the other maps do not show this power-law behaviour clearly, they are also more noisy datasets. We conjecture that some kind of Zipf's law may be a more general feature of these kinds of plots.

It was shown \cite{dingle2020generic} that the sum of the probability mass of outputs with $\log_2$ probability that is at least $\Delta$ bits below the upper bound is no more than $2^{-\Delta+1+O(1)}$. This implies that if there is decay in probabilities, it should be at least exponential. However this does not explain why there is, for all maps, a roughly smooth exponential decay to the same lowest probability. This remains and open question for future work.

\subsection{Is there a bias against changes in natural data?}
We saw that map-specific idiosyncrasies can bias output probabilities via patterns which are hard to make for the map but are not in themselves very complex. But are there any general trends which are likely to be common across different real world maps? The AIT coding theorem in Eq.\ (\ref{eq:CD}) and the related upper bound in Eq.\ (\ref{eq:SB}) are based on fundamental information content of patterns assumed to be generated by UTMs, but in the natural world many patterns may not be physically easy to make. We suspect there is a disconnect between simple patterns with low information content, and those which are easy to make by real world systems. As an example, the patterns $x=$00000001111111 and $y=$01010101010101 have similar complexity from an information perspective, eg they have the same entropy and same size. But it may be that oscillations of $y$ are `harder' to maintain in the physical world than the one simple change in $x$, because change is often expensive in terms of energy or mechanics (with notable exceptions like pendulums, but even they can be easily upset). A more extreme example is the Champernowne's number 0.123456789101112.... which again is a very simple in terms of information content, but it is hard to see how any natural system in the physical world could make such a pattern. Potentially these kinds of patterns may be LKLP in a range of different computable, physically relevant maps.

As a first brief investigation into whether change is `hard' to generate for natural systems and hence associated to LKLP outputs, we study just the frequency of 0/1 changes in outputs. By ``changes'' we mean the number of times a 1 is followed by a 0, or a 0 followed by a 1, eg the string 00111010 has four changes, and the string 0000 has no changes. We ask: for two strings of the same complexity value $k$, does the probability of an output decrease with increasing numbers of 0/1 changes? 

Figure \ref{fig:changes} shows plots for the number of changes vs probability for each map. To aid illustration in the figure, for each map we chose just a single complexity value $k$, which was the second lowest of the output complexity value. The reason for this choice is that LKLP behaviour is most pronounced for very low complexities. For the FST, time series, and RNA there is a clear trend showing an exponential decay in probability with increasing number of 0/1 changes. The polynomial case does not show a trend, but on the other hand there are only two value (2 and 3 changes) which makes observing a trend difficult. 

The linear correlation and $p-$values for the trends are given here in the format of complexity value, correlation coefficient, and $p-$value. For brevity, values are only reported if either the $p$-value was statistically significant (taken as $<0.05$) or the correlation coefficient $r$ was non-trivial, taken here as $r<-0.4$. 
FST: $k=$14, $r$=-0.95, $p$-value=0.01;  $k=$17, $r$=-0.49, $p$-value=3e-05. Time series: $k=$12, $r$=-0.80, $p$-value=3e-08; $k=$14, $r$=-0.44, $p$-value=1e-05; $k=$16, $r$=-0.42, $p$-value=1e-14; $k=$18, $r$=-0.46, $p$-value=1e-50. RNA: $k=$19, $r$=-0.82, $p$-value=0.002; $k=$22, $r$=-0.84, $p$-value=1e-06; $k=$24, $r$=-0.46, $p$-value=3e-18; $k=$26, $r$=-0.46, $p$-value= 6e-38; $k=$29, $r$=-0.45, $p$-value=2e-168.
Polynomial: $k=$16, $r$=-0.95,  $p$-value= 6e-51;  $k=$22, $r$=-0.70,  $p$-value= 6e-321. 

We conclude that a bias for or against changes is a simple mechanism that can quickly lead to exponentially large variation in probabilities, for outputs with the same informational complexity. Further, we tentatively conjecture that in the natural data the bias is more likely to be against changes.

\subsection{Bias for 1s or 0s}
Another simple mechanism which might conceivably strongly affect probabilities is a bias for or against 1's or 0's. In terms of information theory, 1's and 0's are given equal weighting, and all common complexity measures would assign equal complexity to a string and to the same string but with the 0's and 1's flipped, eg 010110 flipped to become 101001. Despite this, it is easy to imagine that a natural system may have a bias for 1's or 0's, and hence the symmetry in information between a string and its flipped counterpart would not extend to a similarity in the probabilities of the two strings (Cf.\ \cite{burkat2022note}). Further, if a discretisation such as converting a real-valued time series into a binary string was performed, then how exactly this was done could bias for 0's or 1's. 

We investigate this potential bias in the four maps described above. Figure \ref{fig:numb_ones} shows the scatter plots for each map, with the number of 1's in a string vs the probability of the string. Again here just one complexity value $k$ for each map is used; if there was no LKLP behaviour, then all strings would have the same probability. The FST example in Figure \ref{fig:numb_ones}(a) has clear bias against 1's, with exponentially decaying probability with more and more 1's. The time series and polynomial data do not display a linear trend, but rather large and small fractions of 1's are associated with lower probabilities, which for the time series can be expected from the way the discretisation was done. The reason is less clear for the polynomial example. The RNA example has an overall bias against 1's, but it is not quite a linear trend. Based on these few examples, there is no obvious common relation between the number of 1's in a string and its probability.

The linear correlation and $p-$value are given now, but only for correlations that were both non-trivial and statistically significant. FST: $k$=14, $r$=-0.95, $p$-value=0.01;  $k=$17, $r$=-0.50, $p$-value=2e-05;  $k=$22, $r$=-0.41, $p$-value=4e-65. Time series: no linear relations. RNA: $k$=19, $r$=-0.82, $p$-value=0.002; $k$=22, $r$=-0.84, $p$-value=1e-06; $k$=24, $r$=-0.48, $p$-value=6e-20; $k$=26, $r$=-0.50, $p$-value=6e-45; $k$=29,  $r$=-0.47, $p$-value=2e-189. Polynomial: No linear correlations.

It seems that exponentially varying probability, unrelated to complexity, can easily arise in natural systems from biases for 1's or 0's and/or changes and oscillations. This helps to understand LKLP behaviour, but a full understanding of the ubiquitous `triangle' scatter plots we see arising from SB remains to be determined.

\section{Applying algorithmic probability}

\subsection{Systems effectively equivalent to a UTM}

Algorithmic probability has been applied or invoked in many different areas of science and mathematics, where implicitly or explicitly the original coding theorem involving a UTM is assumed. No physical computational system can actualise a UTM in the strictest sense, because a UTM requires infinite computational resources, including infinite memory and computation time, while this is not possible in practice. Nonetheless, if a system is at least Turing complete, and if a large memory resource is available which can be expanded as needed, and `long' computations are permitted, then the system can be considered effectively equivalent to a UTM. 

We have seen that for computable maps, low Kolmogorov complexity, low probability (LKLP) outputs are ubiquitous and their existence can be understood from a number of angles. While the original coding theorem in Eq.\ (\ref{eq:CD}) states a direct connection between Kolmogorov complexity and probability, in applications of algorithmic probability to real world scenarios, if it is known that the system under consideration is not effectively equivalent to a UTM, then LKLP outputs should be expected. %Therefore whether or not a system/map of interest has this equivalence property will be an important question in relation to whether LKLP behaviour is observed, and will be a main factor determining whether LKLP outputs are seen. 

Additionally, even if a system has this effective equivalence property and hence has the ability to implement arbitrary algorithms, another issue is whether or not it is reasonable to assume that the system is processing purely random input programs (hence any conceivable algorithm might be implemented), or instead is merely implementing some random inputs to some computable functions for which LKLP outputs might be expected. (Indeed, whether the input programs being processed are reasonably modelled as random at all is another question.) For example, a desktop computer with a Turing complete language compiler like Python is effectively equivalent to a UTM, but if the computer is used to run RNA structure prediction with random RNA sequences, then we can still expect that the outputs we are observing will show LKLP behaviour like in Figure \ref{fig:SB}(c). In this scenario, the UTM-equivalent computational capacity of the underlying system (ie the desktop computer) does not preclude LKLP outputs, and is in a sense irrelevant. Even though the computer is effectively a UTM, we do not expect to see patterns like the algorithmically simple digits of $\pi$ (or perhaps even some non-pseudo random patterns) appearing with high probability, because the programs sampled are not relevant to that type of pattern. 

\subsection{Two questions for physical systems}

The preceding discussion suggests two questions in reference to when we expect LKLP outputs:  Is the physical system of interest likely to be effectively equivalent to a UTM? If effectively UTM equivalent, is it reasonable to assume that the physical system of interest is running random programs for which all manner of algorithms might be computed, or instead computing only a small fraction of computable algorithms?

Regarding the first question, the physical universe does support universal computation, because it can (and is) performed within the universe. Moreover, Wolfram (Chapter 12 of \cite{wolfram2002new}) has proposed the Principle of Computational Equivalence which states that systems in nature which are not obviously simple, eg the weather, have maximally possible computational power, implying that many or even most natural systems operate effectively equivalent to a UTM. Other work on undecidability in physics \cite{moore1990unpredictability,svozil1993randomness,cubitt2015undecidability,aguirre2021undecidability} and the computational capacity of physical world \cite{aaronson2005guest,zenil2013computable,lloyd2006programming} may tend to support the possibility of high-level computation in the natural world. Despite these points, the Principle has not been proven to hold and it is not clear that it does actually hold very commonly in nature. 

Regarding the second question, there are many examples in nature of maps which are clearly not performing arbitrary algorithms, but instead only performing a narrow set of computable functions. In biology, the mapping from DNA sequence programs to biological forms can often be modelled with computable mathematical functions, and hence we can expect LKLP behaviour. Another example: the time series pattern describing the daily temperature in London, UK, over the year has been `computed' from a variety of input factors such as global air currents and cloud movements. But this computation is described by computable mathematical functions (and some uncomputable random noise), and hence we can expect LKLP behaviour if there is any SB in this system. It may be that the weather system could in principe perform arbitrary computations, but it is far from clear that it is actually doing this, and hence LKLP behaviour might be expected. 

Because computable maps abound in nature, and UTM-equivalent systems are not obviously very common, and even if they are, do not commonly implement the full range of possible algorithms, we suggest that LKLP behaviour should be the default assumption in applications of algorithmic probability in real-world settings. We now highlight a selection of  applications of algorithmic probability and see how or if LKLP behaviour is relevant.

%machine learning \cite{valle2018deep,hernandez2021algorithmic,legg2013approximation}, and psychology \cite{gauvrit2014algorithmic}, to name a few examples.

\subsection{Explaining symmetry and simplicity in biology}
It has recently been argued \cite{johnston2022symmetry} that a significant factor causing the symmetry and simplicity in many biological forms (eg in large symmetric biomolecules, petal arrangements in flowers) is due to simplicity bias described by the upper bound in Eq.\ (\ref{eq:SB}). This application of algorithmic probability is unproblematic, because even in the presence of LKLP outputs, there remains a bias towards simpler outputs.

\subsection{Time series prediction}

The authors of ref.\ \cite{dingle2022note} invoked algorithmic probability and the upper bound in Eq.\ (\ref{eq:SB}) to make predictions about natural time series data taken from the World Bank Open Data project (the example time series data set described above was taken from their study). They proposed to make a kind of `forecasting without historical data'. In this prediction context, LKLP outputs do cause a problem for prediction, because although the predicted upper bound closely matches the upper bound of the data, the presence of LKLP outputs imply that many series extrapolations will be predicted to have high probability (on account of being simple) but in fact have low probability. This amounts to a significant weakness in the prediction due to LKLP outputs. A similar challenge to prediction accuracy may also affect the prediction-by-compression of time series studied by Ryabko et al \cite{ryabko2016compression}.

\subsection{What will I see next?}

%%Given that this single question appears in so many in-
%stances in different  fields --- could there be a single, unified
%approach or theory that answers this question in all contexts
%uniformly?
%
%The goal of this work is to provide a proof of principle
%that we can indeed have a theory2 of this kind | one that
%is simple, rigorous, and well-motivated.

 M\"uller \cite{muller2020law} has recently proposed a novel application of algorithmic probability as a framework for addressing fundamental problems in theoretical physics, such as why the universe obeys simple laws. He points out that many problems in science and philosophy reduce to the question ``\emph{What will I see next?}'', and suggests that this single unified approach to framing questions about the world also has a single unified approach to answering them in all contexts uniformly, namely (conditional) algorithmic probability. To see how, we briefly recap some relevant theory \cite{hutter2007algorithmic}: the continuous universal a priori probability $M(x)$ is defined as 
\begin{equation}
M(x) = \sum_{p:V(p)=x*} 2^{-|p|}
\end{equation}
where $p$ is a program that runs on some UTM $V$ and produces an output string which starts with $x$ and then continues (perhaps never halting), denoted by $x*$. So $M(x)$ is the probability that an infinite string begins with $x$.  Now, $M(y|x):=M(xy)/M(x)$ provides a general method for predicting the extrapolation $y$ given a history $x$. Further, Solomonoff showed that even if $\mu$ is some computable (eg real world) probability distribution over binary strings, then the conditional probability prediction problem $\mu(y|x)$ can be estimated by $M(y|x)$, with $\mu$-probability 1 if $|x|\rightarrow\infty$ and $\mu$ is low complexity computable function \cite{solomonoff1978complexity}. %Loosely, this intuitively reasonable framework says that in a very broad setting, predicting the future value of some function can be achieved by looking at which future pattern most `naturally' completes the pattern in the historical data.  The very general nature of this prediction method, and the fact that it should work even if $\mu$ is unknown, is quite striking and invites broad applications. 
We point out that predictions based on $M(y|x)$ may be very inaccurate, ie $M(y|x)\gg \mu(y|x)$, if $y$ may is a LKLP pattern, however.

Returning to the unified approach to framing questions about the world, M\"uller takes as a starting point the assumption that there is only the state of the observer (which is not fundamentally embedded into anything), and then postulates that what happens next to that observer is dictated by algorithmic probability. He goes on to show that this looks to the observer as if he/she was embedded in some computable probabilistic world.  Finally, assuming there is an external world with computable laws within which the observer is embedded, then it is proposed that algorithmic probability gives approximately correct predictions for what the observer sees next, as indicated by the close quantitative relation between $\mu(y|x)$ and $M(y|x)$.

%Because the physical world is governed by various physics laws, which are computable mathematical equations, then it has been argued \cite{muller2020law} that this form of induction should also apply in the natural physical world which we observe and interact with. 

Do LKLP outputs impact the stated goal to address a unified question in a unified manner? Given that algorithmic probability was directly postulated \cite{muller2020law} to be the best way to approach the question of  ``\emph{What will I see next?}'', it is perhaps inappropriate to consider whether it is reasonable to assume the presence of a UTM. Nonetheless it is worth noting that LKLP outputs may be relevant if the manner in which the state of the observer is updated is not in fact reasonably modelled as the result of (random) programs fed into a system which is effectively equivalent to a UTM, but instead as outputs from computable maps.

%To address this we need to consider whether the manner in which the state of the observer is updated is reasonably modelled as the result of (random) programs fed into a system which is effectively equivalent to a UTM, or if instead results from computable maps. It is not entirely clear Hence we suggest that LKLP behaviour is relevant the invoking algorithmic probability to address the question ``\emph{What will I see next?}'' \cite{muller2020law} and this research program may benefit from incorporating LKLP behaviour, for which possibly . 

M\"uller also suggested that algorithmic probability helps to explain why we see a ``simple'' world and physics laws.  Even with LKLP behaviour, this argument would still hold, because the simplicity bias upper bound still favours simple (compressible) outcomes and extrapolations of historical patterns. 

Somewhat related to the preceding, Lloyd \cite{lloyd2010computational} has argued that quantum fluctuations act a random programs which are computed by the universe, to produce complexity in the universe. Lloyd's argument (which invokes AIT) explicitly proposes that the universe acts as a computer; it would be interesting to consider the implications (if any) for LKLP outputs this perspective on the physical universe.

%This implies a very simple and at the same time surprising
%consequence: Solomonoff induction can be used to make successful predictions in our physical world — predictions
%that must agree with those of our best physical theories (or
%even better future theories) if their regime of applicability is
%the regime of data collection. Solomonoff induction will automatically
%“discover” the probabilistic laws of nature that
%we already have (such as quantum theory), or possible future
%ones. In some sense, Solomonoff induction can thus be
%seen as a formal analogue of the scientific method itself.\\

%Solomonoff induction works in the context of computable,
%probabilistic processes — and there is one particularly relevant
%process of this kind: our physical world, as it presents
%itself to the observations of a physicist. Given data on initial
%conditions of some physical system, we can in principle
%write a computer program that simulates the laws of physics
%as we know them and produces predictions for all observations
%that we may perform on the system at later times.\\

%This inequality tells us that transitions to those xy tend
%to be preferred which are “more natural continuations” of
%the previous state x. That is, if xy has a short description\\

%"it explains in some sense ``why'' we see a simple, computable, probabilistic external world"\\

\subsection{Universal gambling}

A universal gambling scheme which considers specific individual outcomes (and explicitly was based on Kolmogorov complexity) was introduced by Cover \cite{cover1974universal} in 1974, in which it was suggested that an investment portfolio should be constructed while respecting probability predictions essentially similar to algorithmic probability. In more detail, an investor having observed a binary string financial time series $x_1x_2\dots x_i$ for some index $i$, in predicting whether the next bit will be a 0, the formula Cover suggested was $p(0)=2^{l(x_1x_2\dots x_i0)}$/$2^{l(x_1x_2\dots x_i)}$, and $p(1)$ is similarly defined. In this formula $l(z)$ measures the minimum codelength for some string $z$, which is fundamentally the same notion as the Kolmogorov complexity of $z$, which appears in the coding theorem.

It seems reasonable to assume that the natural context for applying this universal gambling strategy, that is time series in financial markets, are not the result of UTMs but rather some computable processes. Hence it seems likely that LKLP behaviour would be a challenge for the strategy, due to the possibility that for example $x_1x_2\dots x_i0$ might be simpler than $x_1x_2\dots x_i1$ but due to LKLP behaviour, the latter is more likely. We are not saying that the gambling scheme is invalid, just that its efficacy would be reduced somewhat in the presence of LKLP outputs.

\subsection{Solomonoff induction and Occam's razor}

Occam's razor is a fundamental principle of scientific reasoning, philosophy, and model selection \cite{hansen2001model} stating that simpler explanations or models should take preference over more complex ones \cite{sep-simplicity}. Despite its wide application and common sense appeal, a formal grounding for this principle has been a challenge for philosophy. Solomonoff introduced the idea of algorithmic probability in the 1960's \cite{li2008introduction,solomonoff2009algorithmic} as part of an investigation into induction, and it has been argued that his formal method of induction has solved the long-standing philosophical problem of why simpler explanations should be preferred \cite{vallinder2012solomonoff,kirchherr1997miraculous,rathmanner2011philosophical}. Solomonoff's basic argument is that because simple hypotheses/explanations are a priori more likely to appear from a random program running on a UTM, then given some observed data the simplest hypothesis/explanation should have the highest probability and hence should be preferred (assuming it explains some observed data as well as another competing hypothesis). 

%Expanding on this with an example, suppose the sequence 1,4,9,16, 25 was observed, and we wanted to predict the next value by induction, inferring a general rule from specific example data. For simplicity, suppose we wish to choose between two hypotheses $H_1$ and $H_2$ about the process which generated the sequence, and hence infer the next value  via the value predicted by the chosen hypothesis. Our hypotheses $H_1$ and $H_2$ are given by:
%\begin{center}
% $H_1$: $n^2$ for $n\geq 1$
%\end{center}
%and 
%\begin{center}
% $H_2$: ordered unique solutions to $f(x)=0$, \\repeated indefinitely, where 
% 
% \vspace{0.3cm}
% $f(x)=(n-1)^3(n-4)^{83}(n-9)^4(n-16)^{34}(n-25)^{291}(n-472)^{582}$
%\end{center}
%
%Both $H_1$ and $H_2$ would yield the values 1,4,9,16,25 and so they both agree with the data, but they make different predictions about the next value (36 and 472, respectively). According to algorithmic probability $H_1$ has a higher a priori probability of existing in the world because it is simpler than $H_2$, the latter requiring far more bits to specify in precise detail. Occam's razor would also prefer $H_1$ due to `simplicity', but what simplicity means and how to quantify it are gaps which Solomonoff induction fills in.

Does LKLP behaviour affect the applicability of Solomonoff induction to the real world? If we assume that the contexts within which induction is to be made --- presumably the physical world --- result from random programs fed into a UTM, then LKLP do not even exist. However, as discussed above it seems reasonable to assume that the physical world is the result of computable physical laws and therefore we can expect (many) LKLP outputs.  Solomonoff induction is premised on the fact that a given simple pattern (hypothesis) is a priori more likely than some other given more complex one, and interestingly this property still holds albeit in a weaker sense, even with LKLP outputs. 

Recall that we saw above (Section \ref{whichishigher}) when predicting whether $x$ or $y$ is more likely, that for both probability-weighted and uniform sampling simpler strings tend to have higher probability. However, because uniform sampling the prediction accuracy is low, eg 63\% for uniform sampling with the FST, this suggests that while Solomonoff induction is still valid in LKLP settings, it is less likely to to lead to correct induction as compared to in the UTM setting for which complexity and probability is much more closely connected. Therefore a challenge to applying Solomonoff induction in the real world is faced because when weighing up two hypotheses $H_1$ and $H_2$ that both explain the observed data, $H_1$ may be simpler than $H_2$, but perhaps $H_1$ is a LKLP output and hence much less likely than $H_2$. By extension, there is a challenge to justifying Occam's razor in the real physical world ---  it is still valid even with LKLP, but the argument for preferring simpler hypotheses is somewhat weakened.

Relatedly, Hutter has formed a universal theory of artificial intelligence based on algorithmic probability and Solomonoff induction \cite{hutter2004universal} (see also Legg and Veness \cite{legg2011approximation} for a numerical implementation of the theory). The implications of LKLP outputs for this research project will be similar to those for Occam's razor, namely if the intelligent agent is making observations resulting from an environment that is known to generated by a computable map, then it is still true that often a given simpler hypothesis is a priori more likely than some given complex hypothesis, but not that rarely the reverse is true. Hence in the computable setting it may be that this form of induction is less likely to to lead to correct predictions as compared to the UTM setting.

\subsection{DNN generalisation}

Another recent and important application of simplicity bias is in machine learning, where it has been argued \cite{valle2018deep} that the surprising generalisation ability of deep neural networks (DNN) is due in part to the fact that DNN are biased towards simple functions (by invoking Eq.\ (\ref{eq:SB})), and natural functions are also biased towards simple functions, so the problem of learning functions is significantly easier than it would be if there was no simplicity bias in functions. Because LKLP behaviour is also found in DNN \cite{valle2018deep}, and LKLP has been observed in so many other natural settings, this raises a question: does LKLP behaviour create a challenge for this invocation of algorithmic probability? One way to look at this is to ask, do the simple functions in nature coincide with the simple functions towards which the DNN are biased? Without LKLP outputs in either DNN or nature, they would automatically coincide, but this is no longer automatic given that LKLP outputs exist. If the natural functions are highly probable simple functions but different to the highly probable functions that DNN produce (or at least not a subset of these functions) then the argument for why DNN generalise is weaker. On the other hand, if highly probable natural functions do coincide (or are at least a subset) of the highly probable functions generated by DNN then it would be interesting to consider why this is. This question relates to our earlier question regarding whether there might be common patterns to which types of outputs are LKLP: if different maps or systems have completely unrelated LKLP patterns, then an overlap between different systems is less likely as compared to if there are more general typical patterns across different systems.

\subsection{Password guessing}

Although not based on Kolmogorov complexity and Levin's Eq.\ (\ref{eq:CD}), an essentially very similar probability prediction method has been derived by workers in information theory.  Merhav and Feder \cite{merhav1998universal} point out in an influential review of \emph{universal prediction} that $2^{-LZ(x)}$ is a universal probability assignment for prediction, citing the work of ref.\ \cite{plotnik1992upper} and others. In this context, $LZ(x)$ is the Lempel-Ziv compression complexity measure, which is very similar to the measure used in most studies of SB. Merhav and Cohen \cite{merhav2019universal} have recently used this $2^{-LZ(x)}$ universal probability predictor in a cryptography setting, where they suggested that it forms an optimal method for guessing passwords. It would be interesting to investigate any examples of LKLP passwords. If they did occur, then they might represent a source of wasted guesses of the guessing strategy, because the strategy would assign a high probability of $\sim2^{-LZ(x)}$ to some password $x$, which is in fact very rarely used as a password by people.

\section{Discussion}

We have investigated the occurrence of low Kolmogorov complexity, low probability (LKLP) outputs in computable functions with randomly sampled inputs. The central messages are that (a) LKLP outputs have been observed in essentially all maps for which simplicity bias (SB) has been studied; (b) LKLP outputs are expected in computable maps for theoretical reasons; (c) there appears to be some common statistical patterns to the distributions of the LKLP outputs which form a kind of `triangle' shape in probability-complexity plots; and (d) when applying algorithmic probability in real world applications, LKLP outputs should be the default expectation unless there is good reason to expect that the outputs are indeed generated by purely random programs fed into a universal Turing machine (UTM), which we suggest is probably an uncommon scenario in science, engineering, finance, etc. Further, we briefly surveyed some works in which algorithmic probability in some form has been invoked, and discussed some possible implications of LKLP outputs, including studies of a priori predictions, Solomonoff induction and Occam's razor. The main LKLP implication is that the connection between complexity and probability is considerably less strong, as compared to in the original algorithmic information theory (AIT) coding theorem. 

A main motivation for this study is developing theory for improved a priori probability predictions. The AIT coding theorem states that output probabilities can be directly found via the Kolmogorov complexity of the output, rather than, say, using historical frequency statistics to estimate probabilities. In earlier work \cite{dingle2018input} a practical weaker version of this AIT coding theorem was presented, in the form of an upper bound on probabilities, rather than a direct estimate of probabilities. Understanding the causes and nature of LKLP outputs which fall far below the upper bound may help to improve predictions of the probabilities of those outputs, and hence a stronger theory of a priori probability predictions may be laid out. 

Although algorithmic probability was originally formulated in the context of random algorithms/programs that generate outputs via a computer, we stress that algorithmic probability estimates --- especially the upper bound of Eq.\ (\ref{eq:SB}) --- are not limited to what are usually understood as algorithms per se.  Instead these estimates can be applied to a wide range of problems for which output patterns result from mathematical functions with some form of input parameters. For example, the parameters of a large ordinary differential equation system are not usually understood as an `algorithm' for the solution profile, nonetheless the upper bound has been shown to predict the probability of outputs in such systems \cite{dingle2018input,johnston2022symmetry}. Even more distant from a computer running a program, the upper bound has been shown to work well in predicting natural time series patterns \cite{dingle2022note} for which both the notion of program and computer is much less clearly defined.

There has been a lot of discussion in the statistics and philosophy communities regarding how to choose a Bayesian prior, and AIT promises to be one way to address this \cite{hutter2007universal}. Our work here is directly relevant  to making practical implementations of this AIT answer to the Bayesian prior question.

Several open question remain. In general, how can we better understand the causes and nature of LKLP outputs? Given a small sample of outputs, can these be used to predict which outputs are likely to be LKLP, and which are likely to be close to the bound? Why is it that the data points form `triangles' in Figure \ref{fig:SB}  and other simplicity bias studies? Are there common patterns across different system dictating which outputs will be LKLP? How can we best incorporate simplicity bias probability predictions into other probability estimation approaches, such as machine learning?

\newpage

\vspace{0.5cm}
\noindent
{\bf Acknowledgements:} KD acknowledges financial support from the Gulf University for Science and Technology Seed Grant (grant number 234271). We thank Paris Flood, Iain Johnston, Ard Louis, Nora Martin, Christopher Mingard, and Markus M\"uller for valuable discussions and suggestions related to this work.

\vspace{0.5cm}
\noindent
{\bf Data availability:} The data sets generated during and analysed during the
current study are available from the corresponding authors on request.

\vspace{0.5cm}
\noindent
{\bf Author contributions:} MA performed the numerical calculations. KD conceived the study and wrote the paper.

\bibliographystyle{unsrt}
\bibliography{RNArefs} % expects file .bib

\end{document}